\documentclass{iau}
\usepackage{graphicx}

\title[Simulations of the Fermi bubbles] 
{Fermi bubble simulations: \\
black hole feedback in the Milky Way}

\author[M. Ruszkowski, H.-Y. K. Yang \& E. Zweibel]   
{M. Ruszkowski$^1$
H.-Y. K. Yang$^1$
\and  E. Zweibel$^2$}

\affiliation{$^1$Department of Astronomy, University of Michigan, Ann
  Arbor, MI, USA \\ email: {\tt mateuszr@umich.edu, hsyang@umich.edu}\\[\affilskip]
$^2$ Department of Astronomy and Physics, University of Wisconsin–Madison,
Madison, WI, USA \\ email: {\tt zweibel@astro.wisc.edu}}

\pubyear{2014}
\volume{xxx}  
\pagerange{1--4}
\jname{The Galactic Center: Feeding and Feedback in a Normal Galactic Nucleus}
\editors{A.C. Editor, B.D. Editor \& C.E. Editor, eds.}
\begin{document}

\maketitle

\begin{abstract}
The {\it Fermi} gamma-ray telescope discovered a pair of bubbles at
the Galactic center. These structures are spatially-correlated with
the microwave
emission detected by the WMAP and {\it Planck} satellites. 
These bubbles were likely
inflated by a jet launched from the vicinity of a supermassive
black hole in the Galactic center. 
Using MHD simulations, which self-consistently
include interactions between cosmic rays and magnetic fields, 
we build models of the supersonic jet propagation,
cosmic ray transport, and the magnetic field amplification within the
{\it Fermi} bubbles. Our key findings 
are that: (1) 
the synthetic {\it Fermi} gamma-ray and WMAP microwave
spectra based on our simulations are consistent with the observations,
suggesting that a single population of cosmic ray leptons may
simultaneously explain the emission across a range of photon energies;
(2) the model fits the observed centrally-peaked microwave emission 
if a second, more recent,  
pair of jets embedded in the {\it Fermi} bubbles is included in the
model. This 
is consistent with the 
observationally-based suggestion made by \cite{sufinkbeiner};
(3) the radio emission from the bubbles is expected to be strongly
polarized due to the relatively high level of field ordering caused by
elongated turbulent vortices. This effect is caused by the interaction of
the shocks driven by the jets with the preexisting interstellar
medium turbulence; (4) a layer of enhanced rotation measure in the
shock-compressed region could exist in the bubble vicinity but the
level of this enhancement depends on the details of the magnetic
topology. 
\end{abstract}

\firstsection 
\section{Introduction}
The {\it Fermi} bubbles are large--scale bipolar outflow bubbles in the
Galaxy that were recently detected with the {\it Fermi} observatory 
(\cite{su}). 
They are a unique example of supermassive black hole feedback in our
own ``backyard'' rather than in a distant extragalactic source
(for alternative interpretation see, e.g., \cite{crocker}; 
contributions by Lacki and Dogiel in these proceedings). The {\it
  Fermi} bubbles have received a lot of attention in the literature 
since their discovery and are one of the most exciting findings from
the NASA's {\it Fermi} mission. The bubbles are characterized by a 
flat gamma-ray intensity distribution, hard spectrum extending up to $\sim$100 GeV, and sharp
gamma-ray edges separating the bubbles and ambient interstellar medium.
The bubbles are also visible in the WMAP and {\it Planck} data. This
emission is referred to as the WMAP haze 
(\cite{finkbeiner}) or Planck haze (\cite{PlanckCollab}). This microwave haze
emission is centrally-peaked, i.e., its intensity is strongest near
the Galactic center.

\section{MHD simulations of the {\it Fermi} bubbles}
\begin{figure}[t]
\begin{center}
\includegraphics[width=0.925\textwidth]{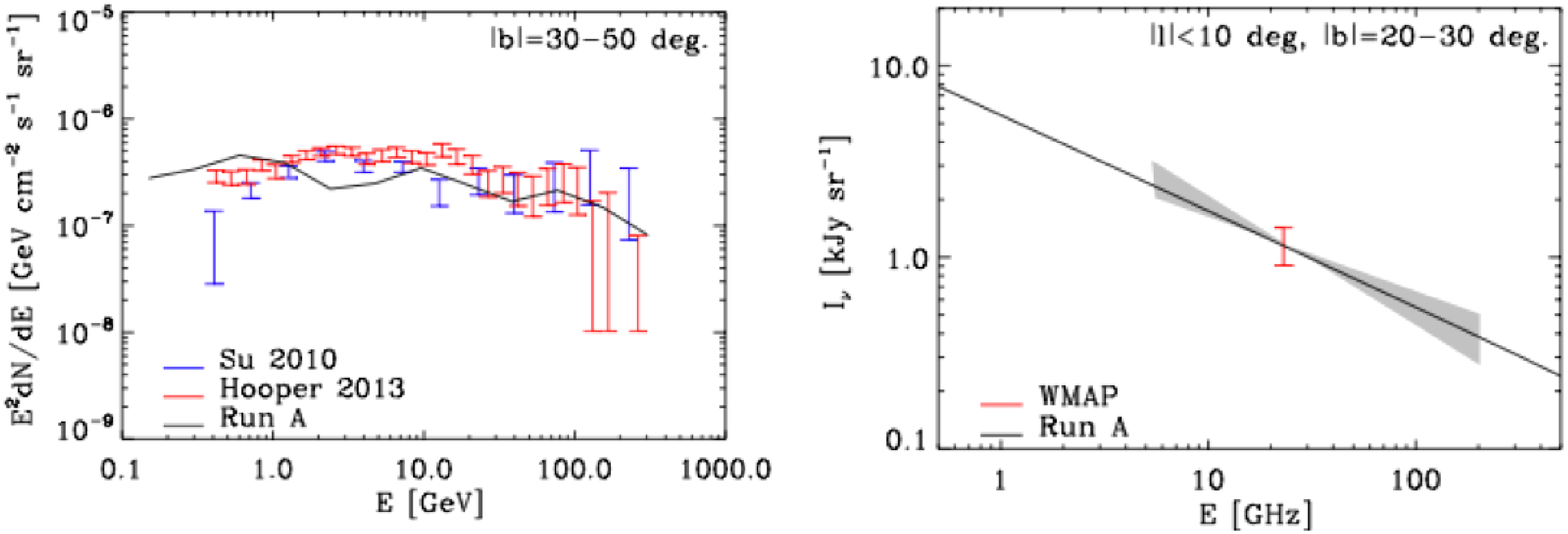} 
 \caption{Left: Gamma-ray (left) and microwave spectra of the
   {\it Fermi} bubbles. Solid lines represent model predictions. $l$
   and $b$ denote Galactic longitude and latitude, respectively.} 
   \label{fig1}
\end{center}
\end{figure}

Using 3D MHD simulations that include cosmic ray physics, we
predict observational signatures of the {\it Fermi}
bubbles in the leptonic AGN jet scenario (\cite{yrz13,y12}). In this model, the gamma-ray
emission is due to the inverse Compton scattering of the cosmic
microwave background, and the interstellar radiation field, by the cosmic
ray electrons inside the jet-inflated bubbles, while the WMAP haze emission is due to the
synchrotron emission produced by the same population of cosmic rays
interacting with the magnetic fields inside the bubbles. \\
\indent
One important aspect of the model is that the cosmic rays tend to
accumulate close to the bubble surface. That is, the intrinsic spatial
distribution of the cosmic rays is edge-brightened. This naturally leads
to a flat distribution of the sky-projected gamma-ray emission of the
bubbles, which is in agreement with observations. 
An appealing feature of this model is that the inflation of the 
bubbles is so rapid that the age of the bubbles is shorter than the
cooling time of cosmic ray electrons. This ensures that the cosmic ray
cooling losses do not significantly alter the spectrum, which remains
hard up to energies as high as $\sim$100 GeV as observed (see Fig.\,\ref{fig1}).

If we assume an observationally motivated distribution of the 
ambient Galactic magnetic fields 
(\cite{strong}), we can combine it with the simulated cosmic ray distribution
to match the observed slope and normalization of the WMAP synchrotron
emission (see Fig.\,\ref{fig1}). We stress that this {\it match between the
simulated and observed gamma-ray and microwave spectra} can be obtained
by invoking a {\it single population} of cosmic ray leptons.
Thus, the model
combined with observations allows us to essentially {\it look inside the Fermi bubbles.}
Interestingly, in the leptonic model, the cosmic rays contribute relatively little pressure compared to
the pressure of thermal gas inside the bubbles. Finkbeiner (these
proceedings) reported on a 
cutoff in the gamma-ray spectrum beyond $\sim$100 GeV. This observation will help to put additional constraints
on the cosmic ray energy spectra, cooling/heating mechanisms,
and the bubble dynamics early in its evolution.\\
\begin{figure}[t]
\begin{center}
\includegraphics[width=0.925\textwidth]{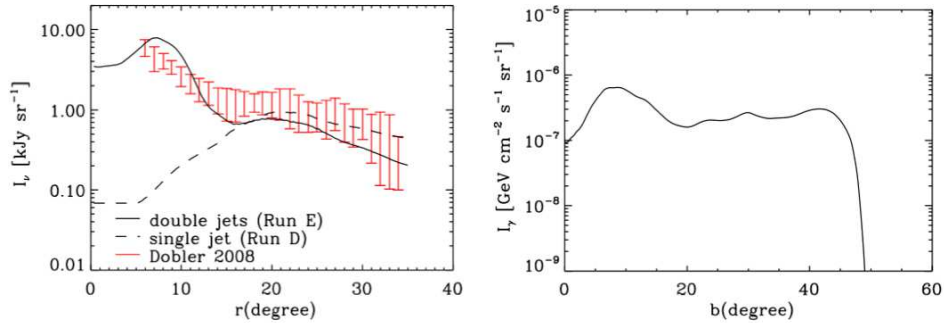} 
 \caption{Microwave (left) and gamma-ray intensity
   profiles.} 
   \label{fig2}
\end{center}
\end{figure}
\indent
{\underline{\it Magnetic field amplification inside the Fermi
    bubbles}} As the bubbles expand, the ambient ISM magnetic fields drape
around the bubbles. 
Assuming cosmic ray diffusion is anisotropic (\cite{y12}), this
prevents the cosmic rays from
readily diffusing out of the bubbles, and thus accounts for their sharpness
in the gamma-ray emission.
While the draping mechanism helps to explain this aspect of the
observations, it reduces the strength of magnetic fields inside
the bubbles. That is, the bubbles are effectively sealed off from the
ambient magnetized ISM and the magnetic fields inside them are reduced
due to adiabatic expansion.  Consequently, the cosmic rays inside the
bubbles are in contact with relatively weak magnetic fields and the
predicted microwave emission is too weak compared to observations.\\
\indent
We suggest that one solution to 
this problem is to allow for partial draping of the ambient field. In
order to study this possibility, 
we consider tangled ambient magnetic fields characterized by a range of
coherence lengths. Our fiducial model consist of the global tangled
field superimposed on a more tangled field component closer to the disk plane.
We find that decreasing the coherence length of the latter B-field
component does lead to partial magnetic draping of the bubbles while not
violating the observational constraints on the sharpness of the
bubble-ISM interface seen in the gamma-ray emission. At the same time, the
reduced coherence length of the ambient field allows for some
mixing-in 
of the ambient field into the bubbles. 
We demonstrated that, as this coherence length is reduced, 
{\it the field strength inside the bubbles asymptotically tends to
a value comparable to
that of the ambient medium}. This close relationship between the
ambient ISM field and the internal bubble field is ultimately caused
by the interaction of the shock wave associated with the expanding
bubble with the preexisting turbulent and magnetized ambient ISM. The
shock wave propagating ahead of the bubble increases the level of
vorticity in the post-shock ISM (\cite{larsson}), which in turn amplifies the magnetic
field in the shocked gas. 
The amplification rate increases as the coherence length decreases. This amplified field then mixes into the
bubble though the bubble-ISM contact discontinuity. \\
\indent
As mentioned above, when the cosmic
rays in the bubbles are in contact with magnetic field of this
magnitude, then the simulated microwave spectrum may be reconciled
with the WMAP observations.\\
\indent
{\underline{\it Intermittent activity of Sgr A$^{*}$-- second jet within the Fermi bubbles}}
As the coherence length of ambient field is reduced, the magnetic
fields inside the bubbles increase in strength. When these fields
reach the magnitude required to account for the level of microwave
emission at low galactic latitudes, the plasma $\beta$ approaches
unity, i.e., the fields become dynamically important in these
regions. Consequently, 
the magnetic pressure forces inside
the bubbles act to redistribute the cosmic ray
energy density to higher latitudes. At this point, we are again faced with the
problem that sufficiently strong magnetic fields and cosmic rays are not in contact
with each other. In order to alleviate this problem, we postulated the
existence of a second, more recent AGN activity, and experimented
with replenishing cosmic rays by a weaker, {\it second pair of jets}. These
experiments were successful in explaining the microwave intensity
profiles of the WMAP haze while remaining fully consistent with the
spatial flatness of the gamma-ray emission from the {\it Fermi}
bubbles (see Fig.\,\ref{fig2}).  Based on the observational
evidence, \cite{sufinkbeiner} showed that a better match to the
gamma-ray morphology of the bubbles can indeed be obtained 
if an additional pair of jets, anti-symmetric with respect to the disk
plane, is included in the fitting procedure (see also invited talk by Finkbeiner in these
proceedings).  The {\it Fermi} bubbles, including such 
second pair of internal jets, bear an interesting morphological similarity to Cen
A, though this object hosts a substantially more massive central
black hole than our Galaxy. Interestingly, \cite{li} recently presented 
evidence for a sub-parsec X-ray jet in the Galactic center (see Li's
poster contribution in these proceedings). This cumulative evidence strongly
suggests that the black hole in Sgr A$^{*}$ experienced episodes of
past activity. \\
\indent
{\underline{\it Radio polarization and rotation measure}}
Our model naturally lends itself to predicting radio signatures of the
{\it Fermi} bubbles.
The relatively high level of field ordering inside
the bubbles that was mentioned above leads to {\it high intrinsic linear
polarization fractions}. The bubble fields tend to be stretched along
the jet propagation direction (see radial enhancements in the magnetic field in Fig.\,\ref{fig3}). 
\begin{figure}[t]
\begin{center}
\includegraphics[width=0.9\textwidth]{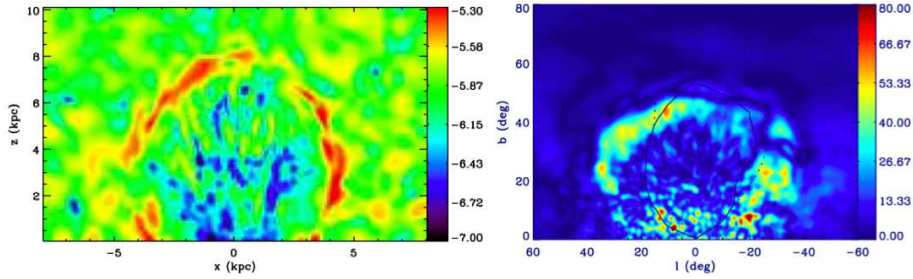} 
 \caption{Slice through magnetic field strength distribution (left)
   and rotation measure.}
   \label{fig3}
\end{center}
\end{figure}
This is consistent with 
the linearly polarized intensity S-PASS maps (\cite{carretti})
that show high level of polarized emission in features 
approximately aligned with the jet propagation direction. The
polarization fractions predicted by the model are somewhat above those
observed, but this discrepancy could easily be resolved by foreground
depolarization. We also make predictions for the rotation measure  
distribution (see Fig.\,\ref{fig3}). Our simulations suggest that the rotation measure could
be boosted around the bubbles. This enhancement is especially likely near the
bubble-ISM interface but not necessarily perfectly coincident with it.  
While the likely presence of magnetic field fluctuations unaccounted for in our
simulations precludes making any definite
predictions as to the level and precise location of these
enhancements, this effect could be searched for in latitude-binned rotation
measure observations.\\

\noindent
We would like to thank the organizers for an exciting and inspiring
meeting, and Roland Crocker, Doug Finkbeiner, Jerry Ostriker, Ann Mao, Meng Su, Fred
Baganoff, Roman Shcherbakov, and Zhiyuan Li for very stimulating discussions during the workshop.
We acknowledge the support from 
NSF AST 1008454 (PI Ruszkowski), NASA 12-ATP12-0017 (PI Ruszkowski),
and {\it Fermi} GSFC 61252, NSPIRES 12-FERMI12-0012 (PI Yang).

\end{document}